\documentclass[%
 amsmath,amssymb,
 aps,pre,onecolumn,superscriptaddress,floatfix
]{revtex4-2}
\usepackage{sidecap}
\usepackage{graphicx}
\usepackage{caption}
\usepackage{dcolumn}
\usepackage{bm}
\usepackage{mathtools}
\usepackage{amsmath}
\usepackage{chemmacros}
\usepackage{xcolor}
\usepackage[version=4]{mhchem}
\usepackage[export]{adjustbox}

\newcommand{\resub}[1]{\textcolor{black}{#1}}

\begin{document}
\title{Chemical reaction motifs driving non-equilibrium behaviors in phase separating materials}

\author{Dino Osmanovi{\'{c}}}
\email{osmanovic.dino@gmail.com}
\affiliation{Department of Mechanical and Aerospace Engineering, University of California, Los Angeles, Los Angeles 90095, CA, USA}
\author{Elisa Franco} 
\affiliation{Department of Mechanical and Aerospace Engineering, University of California, Los Angeles, Los Angeles 90095, CA, USA}
\affiliation{Department of Bioengineering, University of California, Los Angeles, Los Angeles 90095, CA, USA}
\date{\today}

\begin{abstract}
  
Chemical reactions that couple to systems that phase separate have been implicated in diverse contexts from biology to materials science. However, how a particular set of chemical reactions (chemical reaction network, CRN) would affect the behaviors of a phase separating system is difficult to fully predict theoretically. In this paper, we analyze a mean field theory coupling CRNs to phase separating materials and expound on how the properties of the CRNs affect different classes of non-equilibrium behaviors: \resub{ the emergence of microphase separation or of temporally oscillating patterns. We examine the problem of achieving microphase separated condensates by first considering tractable problems and illustrating the mathematical conditions leading to microphase separation. We then identify CRN motifs that are likely to yield size control by examining randomly generated networks and parameters. By analyzing the probabilities to observe particular states, we define simple design rules of CRNs that lead to desired non-equilibrium behavior. We show that  chemical interactions generating negative feedback facilitate microphase separation, moreover, we demonstrate that the parameters important for the emergence of microphase separation differ for systems with two or four components, due to frustration.} Our results provide guidance toward the design of self-regulating material CRNs and provide instructions to manage the formation, dissolution, and organization of compartments.

\end{abstract}

\maketitle

\section{Introduction}

The ability of mixtures of molecules to separate into distinct compartments has major implications in chemistry, materials science, and biology, and offers intriguing hypotheses on the origin of life~\cite{Hyman2014,brangwynne_2009,Alberti2021}. Within cells, phase separated biomolecular condensates  form in response to internal and external stimuli~\cite{riback2017stress}, operating as microscopic organelles without a membrane that enable precise regulation of cell physiology and chemistry~\cite{frottin2019nucleolus, o2021role}. Amongst the biological questions pertaining phase separation observed in living cells, the interplay between biochemical reactions and phase separation is of particular interest, because it can explain the dynamic nature of biomolecular condensates in the homeostatic cellular environment~\cite{alberti2017phase,Banani2017}. In parallel, the combination of synthetic phase separating systems and chemical reactions is making it possible to design  artificial organelles~\cite{schuster2018controllable,reinkemeier2019designer,Klosin2020} and functional materials with precise spatio-temporal responses~\cite{heckel2021spinodal,gong2022computational,agarwal2022biochemical}.  Chemically active droplets can also be made to self-propel, acting as potential carriers of material~\cite{Michelin2022,Toyota2009}. 

  With mean field theories, it has been possible to observe that chemical reactions fuel ``life-like'' behaviors such as droplet splitting \cite{Browne2010,zwicker_2016}, providing that there are sufficient free energy sources available to keep the system out of equilibrium. Even in the absence of chemical reactions, multicomponent phase separating materials display a multitude of morphologies~\cite{Mao2019a}, which can be tuned through control over the self-attraction of the phase separating materials~\cite{Mao2020}. Prior work has shown that chemical reactions can control condensate properties, with focus on examples that allow for the derivation of exact conditions~\cite{zwicker2022intertwined,kirschbaum2021controlling,longo2022phase}, or that include enzymatic reactions particularly relevant in specific biological processes~\cite{zwicker2014centrosomes,Zwicker2015,zwicker_2016,Cates,weber_2019,Brackley2017}. 

Here we \resub{seek general design rules for chemical reactions that can provide a means to control macroscopic non-equilibrium behaviors of phase separated condensates}. Motivated by the rapid expansion of programmable molecular substrates that enable the synthesis of nearly arbitrary chemical reaction networks (CRNs)~\cite{chen2013programmable,chen2021programmable}, we consider the following general questions: over many different chemical reaction networks \resub{coupled with a phase separating material, are there any broad features that arise in the CRNs themselves that lead to particular features of macroscopic states? Furthermore, if the chemical species have some non-chemical coupling (such as non-specific interactions leading to spatial agglomeration of different species), how do these couplings impact the CRNs that lead to those defined features?} Answers to these questions will provide guidance for the design of chemical reactions that can regulate the properties of individual or multiple condensates~\cite{shrinivas2021phase}.

We focus on two particular features of the condensates. One feature relates to average size of the observed condensates. This question can be examined through a mean field model that combines spatial and chemical interactions among species, and corresponds to the problem of studying linear stability of equilibria~\cite{Cross1993,glotzer1995reaction}. Standard phase separation kinetics, such as in the Cahn-Hilliard model, feature \textit{coarsening}, i.e., the distinct phases continue to grow until the system has been separated into macroscopic compartments (macrophase separation). \resub{We seek to identify general chemical reactions} that can arrest coarsening~\cite{glotzer1995reaction}, and lead to a case in which the stationary state of the system contains many droplets with a defined length scale and the system exhibits microphase separation. The other feature we are interested in is dynamical instability. Are there any CRNs which lead to continual (or at least long lived) dynamical oscillations of the system with phase separation?

To answer these general questions, we analyze a non-equilibrium dynamical theory of phase separation kinetics coupled to an \emph{arbitrary} CRN. Through computations, we explore which features of a CRN are sufficient to produce non-equilibrium phenomena, such as condensate size control~\cite{Zwicker2015} and  oscillatory dynamics. \resub{Introducing such a question poses a theoretical challenge, as the number of parameters involved in the problem proliferates very rapidly with the number of components. We therefore seek simplified ``design rules" that can allow us to gain an intuitive understanding of how different parameters affect different features. We achieve this by analyzing the system over the entire parameter space and seeing whether any strong signals exist in lower dimensional representations of the system that can guide understanding. To that end the paper is organized as follows: In the first section we describe the mathematical model. We then expound upon an analytical example for a system with a small number of components as a description of our method. Subsequently, we analyze numerically the same problem with a larger number of components using random matrices. Next, we introduce real chemical reaction networks and repeat our analysis over structural features of the same. Finally, we extend the same analysis to predictions of continual dynamics. }

\section{Results}

\subsection{Modeling phase separation in the presence of chemical reactions}
We introduce a set of $N$ density fields in space corresponding to the concentrations of different chemical species, $\mathbf c(\mathbf x) =(c_1(\mathbf x),c_2(\mathbf x),\hdots,c_N(\mathbf x))$, where $\mathbf c\in \mathbb{R}^{N}_{+}$, $\mathbf x\in \mathbb{R}^d$. We assume these species are coupled through \textit{conserved dynamics} (which we also call spatial coupling), that affect how the density fields arrange in space while keeping their total concentration constant, and through \textit{non-conserved chemical reactions} that determine how the density fields inter-convert. 
 
\begin{figure}[h]
\begin{center}
\includegraphics[width=\textwidth, frame]{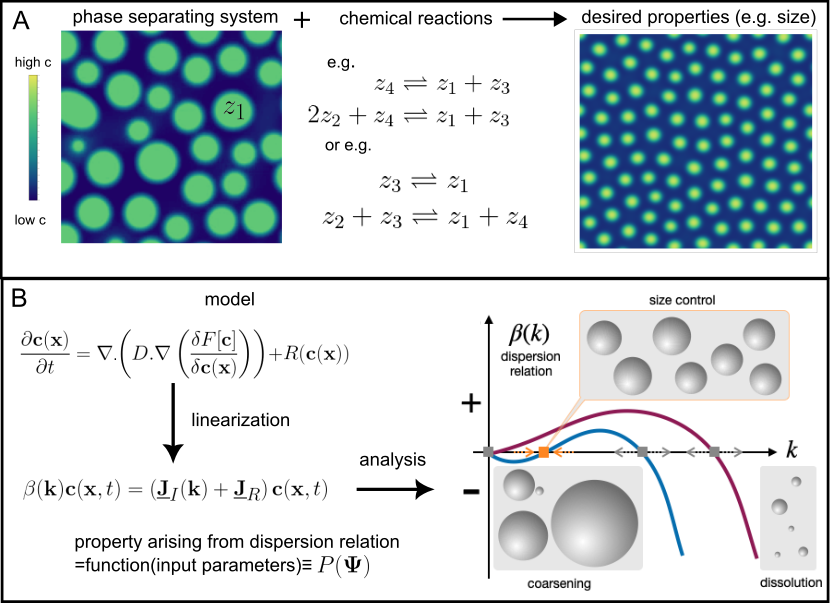}
\caption{\resub{(A) We are considering the interplay of chemical reactions and phase separation in a system in which species $z_1$ phase separates while other species do not. Shown is a Cahn-Hilliard type phase separating system which is then subject to a particular choice of chemical dynamics, which leads to microphase separation (B) Dispersion relations can be obtained from the Jacobians associated with the model. There are two different categories of dispersion relation for phase separating systems: in red we illustrate the case of \resub{macrophase separation}, corresponding to standard phase separation where the phases continue to grow inside until they are macroscopically separated; in blue we illustrate the case of microphase separation, where large droplets (small $|\mathbf{k}|$) and small ones (large $|\mathbf{k}|$) have a negative growth rate, stabilizing droplets to a particular size near the orange intersection point. }}
\label{fig:fig0} %
\end{center}
\end{figure}

To model spatial coupling, we introduce the following equilibrium free energy for the different chemical fields:
\begin{equation} \label{eq:CHF}
F(\mathbf c)=\int \left \{ \mathrm \nu (c_1(\mathbf x)-\rho_1)^2 (c_1(\mathbf x)-\rho_2)^2 + \gamma^2 |\nabla c_1(\mathbf x )|^2  + \frac{1}{2}\mathbf{c}.\boldsymbol\epsilon.\mathbf{c} \right \} d \mathbf x,  
\end{equation}
where $\gamma>0$ is the surface tension,  $\epsilon \in \mathbb{R}^{N\times N}$, and $\nu$ is some energy scale.
This free energy corresponds to a system where component $1$ undergoes phase separation, under Cahn-Hilliard form, to two phases with density $\rho_1$ and $\rho_2$ (Fig.~\ref{fig:fig0}). We will use interchangeably the term ``condensates'' or ``droplets'' to indicate  these phase separated regions. The symmetric matrix $\boldsymbol \epsilon$  introduces some additional spatial coupling between all the fields. Note that because the interaction of $c_1$ with itself is already included in the Cahn Hilliard term, the corresponding matrix component is taken to be $\boldsymbol{\epsilon}_{11}=0$. In principle, we could include higher order couplings of different fields, however for simplicity we truncate the series after $\sim c^2$. We additionally ignore problems related to the total packing fraction of different components and of surface effects of the non-phase separating chemicals (surface tension of components $i\ge 2$). The choice of field $1$ as the phase separating species is merely a matter of convenience, and our results do not depend on this choice.  

The time evolution of this system can be inferred using model B dynamics~\cite{Hohenberg1977}:
\begin{equation} \label{eq:fint}
\frac{\mathrm d \mathbf c(\mathbf x,t)}{\mathrm d t} = \nabla \cdot \left(\boldsymbol{D}.\nabla\frac{\delta F(\mathbf c(\mathbf x,t))}{\delta \mathbf c(\mathbf x,t)}\right)=\mathbf I(\mathbf c(\mathbf x,t)).
\end{equation}

For simplicity we assume the diffusion matrix $\boldsymbol{D}$ is given by $\boldsymbol{D}=\mathcal{D}\boldsymbol{Id}$ where $\mathcal{D}$ is a diffusion constant and $\boldsymbol{Id}$ is the identity matrix. This corresponds to every species having the same homogeneous diffusion coefficient \resub{and molecular mass. We note however, that our model is similar to having cross-diffusion terms with $\boldsymbol \epsilon=0$. It can also be seen that the values along the diagonal of the matrix $\epsilon_{ii}$ where $i>1$ would correspond to rescalings of the diagonal of the diffusion matrix in model B dynamics. Therefore, each term is set to be equal to unity, $\epsilon_{ii}=1, \text{ with } i>1$.} 

\resub{We are left with how to specify the chemical dynamics. Previous literature uses different choices for deriving non-conserved dynamics based on variations of model A dynamics, involving specification of a free energy functional. We do not constrain ourselves into this choice yet, and for now we merely specify that chemical dynamics must obey the following equation: }
\begin{equation}\label{eq:CRN}
    \frac{\mathrm d \mathbf c(\mathbf x,t)}{\mathrm d t} = \mathbf R(\mathbf{c}(\mathbf x,t)),
\end{equation}
where $\mathbf{R}$ are the reaction fluxes generated from the law of mass action by a set of different chemical reactions. We restrict our analysis to chemical reactions which conserve the total mass (no spontaneous creation or destruction of mass). \resub{We keep deliberately general the precise form of $\mathbf{R}$ in equation~\eqref{eq:CRN} for the moment, in order to analyze the generic effects of chemistry on phase separation.}


We finally combine the conserved dynamics (the dynamics that moves the concentration fields around while keeping their total amount constant) and the chemical dynamics (the dynamics that inter-converts concentration fields between one another, thus changing total amounts of individual components) via the following combined equation:
\begin{equation} \label{eq:fum}
\frac{\mathrm d \mathbf c(\mathbf x,t)}{\mathrm d t} = \mathbf I(\mathbf c(\mathbf x,t))+ \mathbf R(\mathbf c(\mathbf x,t))
\end{equation}
This equation is inherently non-equilibrium, as the origins of the two different terms differ. An equilibrium theory would have both terms arising from the functional derivative of the same free energy functional~\cite{Li2020}. We don't focus on this case as we are only interested in non-equilibrium properties resulting from the action of chemistry, as opposed to equilibrium phenomena that can lead to microphase separation (e.g. surfactants). \resub{In contrast to previous studies, such as Li and Cates~\cite{Li2020}, we do not assert a functional to derive the term $R(\mathbf c(\mathbf x,t))$, in fact, for anything other than the simplest form of reactions, there would not be a single unique functional that describes mass-action kinetics in a chemical reaction network. Our approach is instead a specification of the rates at which different reactions proceed rather than from specification of the energies}. This would correspond to a free energy source being constantly consumed to maintain these reactions at the specified rates~\cite{Cates,zwicker2022intertwined}, an approach that was used to study problems such as protein aggregation~\cite{Chan2019}.  We use~\eqref{eq:fum} for reasons of analytical tractability, as a full theory containing phase separating dynamics and chemical reaction networks is beyond the present scope of the manuscript. Our approach assumes that each of the chemical species contains sufficient complexity to have reactions that can change their conserved dynamics with other molecules, either through mechanisms such as allostery or additional steric effects, as previously considered in~\cite{Osmanovic2019}.

We analyse our  model~\eqref{eq:fum} through a linear stability analysis near equilibrium: $\mathbf c(\mathbf x,t) =  \mathbf c_0 + \boldsymbol \delta \exp(i \mathbf k. \mathbf x + \beta t)$, inserting this form into  ~\eqref{eq:fum} allows us to define a dispersion relation $\beta(|\mathbf k|)$(growth rate as a function of wavenumber), where  $\beta(|\mathbf k|)$ is the \resub{(dominant eigenvalue)}. The initial state is chosen by perturbing the system \resub{around the stationary state of the chemical dynamics, $c_0$, as this is the stable homogenuous equilibria}. The dynamics are linearized around $c_0$, by defining the Jacobians: 
\begin{equation}
    J_I(\mathbf k)=\nabla_{\mathbf c(x)} \mathbf I(\mathbf c( \mathbf x,t))\big|_{\mathbf c=\mathbf c_0}, \quad J_R=\nabla_{\mathbf c(x)} \mathbf R(\mathbf c(\mathbf x,t))\big|_{\mathbf c=\mathbf c_0},
\end{equation}

where  $J_I(\mathbf k)$ is the Jacobian matrix of the spatial interaction terms and $J_R$ is the Jacobian matrix of the chemical interactions term. We obtain a generic relationship in $\beta(|\boldsymbol k|)$: 
\begin{equation}\label{Jacobians}
    \beta(|\mathbf k|) \boldsymbol \delta =  (J_I(\mathbf k)+J_R)\cdot \boldsymbol{\delta}.
\end{equation}
where it can be seen that $\beta(|\boldsymbol k|)$ is the eigenvalue of the combined matrix $(J_I(\mathbf k)+J_R)$. We are interested in the occurrence of condensates with finite size (microphase separation). This case arises when $\beta(|\mathbf  k|) < 0$ for small $| \mathbf k | $, $\beta(\mathbf  k) > 0$ for intermediate $|\mathbf k|$, and $\beta(|\mathbf  k|) < 0$ for large $|\mathbf  k|$. We illustrate this with a scalar example in Fig.~\ref{fig:fig0}, where we compare two dispersion relation curves in blue (microphase separation) and in red curve  (macrophase separation). The blue curve indicates that large condensates and small condensates have a negative growth rate, while condensates of intermediate size can have a positive growth rate. This results in the orange crossing point being a stable equilibrium wave number i.e. condensate size is stabilized, and we will also refer to this type of dispersion curve as size-controlled condensation.   \resub{As we only study the appearance of an instability due to an infinitesimal perturbation $\delta$, our analysis is constrained to the spinodal region of the phase diagram, i.e., we do not consider nucleation effects. While it could plausibly be the case that the homogeneous state is stable to small perturbations but not stable globally, we do not study this in this paper.}


\resub{The matrix $J_I(\mathbf k)$ has an exact form that depends on the parameters, however there are two different approaches for generating Jacobians of the chemical reactions $J_R$. In the first approach, we will generate a random matrix for $J_R$, analyze the resulting features of matrix itself, and infer design principles for candidate reaction networks. In the second approach, we will generate random CRNs \cite{VanDerSchaft2011}, derive their Jacobian algebraically, and evaluate whether the \textit{network} can induce microphase separation under a parameter sweep. A major difference between these two approaches is that the second approach yields Jacobians whose entries are correlated, due to the law of mass action. Further, these approaches lead to qualitatively different types of ``design rules" towards the design of chemical networks. While the first approach can tell us about what the mean relationships between species are necessary (i.e., "component $z_i$ should repress component $z_j$"), the second approach can tell us more about the actual networks themselves, which is more useful for experimental design. We shall discuss these two different approaches more in the next section.}


\subsection{Analysis of randomly generated Jacobian matrices}
\resub{In this section we seek to understand how the Jacobian modeling chemical reactions affects the system's capacity to undergo microphase separation. For this purpose, we build a function $P_J$} that takes as its input a Jacobian arising from the chemistry and the Jacobian matrix arising from the interactions, and as its output (for example) produces one of two outcomes, \resub{either microphase separation or macrophase separation} (in the following, we shall ignore the possibility that there is no demixing at all, as it is easier to infer the rules that lead to no phase separation):
\begin{equation}\label{DemixingRule}
    P_J(J_I(\mathbf k),J_R)=
\begin{cases}
1 &\text{if microphase separation,}\\
0 &\text{if macrophase separation}
\end{cases}
\end{equation}
which can be determined through analysis of the dispersion curve. \resub{In other words, given that the entire form of dispersion curves as seen in Fig.~\ref{fig:fig1} is determined by the Jacobians $J_I(\mathbf k)$ and $J_R$, we can associate each pair of matrices with a single number corresponding to whether those Jacobians display microphase or macrophase separated dispersion curves.} This is a deterministic function, in that every unique pair of matrices $J_I(\mathbf k)$ and $J_R$ will map onto the Boolean value. 

\resub{Such a function of matrix arguments is difficult to analyze. It is especially difficult to establish ``design rules" that are apparent from this function. Moreover, while a pair of matrices can be mapped onto a number, there is another problem in that the space of possible matrices is much larger than the space of matrices that arise from consideration of real chemical reaction networks. Therefore, this function gives us ``bare" features that are associated with the Jacobians, but does not tell us what chemical reaction network could give rise to such a Jacobian (we shall analyze this later). We restrict our analysis to  Jacobians $J_R$  that have eigenvalues with negative real part and have a determinant of zero (i.e. are singular). This would correspond to doing a linear perturbation about a stable equilibrium in the equations modeling chemical reactions. It can now be seen that generating a ``real" CRN and then using that as an input into ~\eqref{DemixingRule} is consistent with constraining the distribution of possible Jacobians $J_R$, i.e. a constraint on possible \textit{inputs}. However, the precise form of this constraint cannot be easily evaluated (It correlates all the entries into the matrix). Therefore, it is appropriate to first analyze the function ~\eqref{DemixingRule} over all inputs without these correlations to see what the ``bare" features are (as the function is deterministic), and then perform different sets of measurements over CRNs.  }

\resub{Another issue is that the dimension of arguments to this function is rather large. For example, a two component system would include 4 independent arguments corresponding to all the entries in the matrix $J_R$ as well as the independent parameters in the matrix $J_I(\mathbf k)$ (such as the interaction between components 1 and 2, $\epsilon_{12}$, the surface tension $\gamma$, the diffusion coefficient $\mathcal{D}$ etc.) If our goal is to understand how particular features of the resulting condensates result from a consideration of these parameters, the explosion of parameters as one goes to larger amounts of components impedes understanding. We therefore need a way to be able to establish simple rules that lead to microphase separation in this large space of parameters. Fortunately, such methods have already been established in diverse contexts. However, for completeness, we below sketch what this would entail; for a full discussion, see~\cite{molnar2020interpretable}. }
 
\resub{We first aggregate all the relevant parameters of the problem in a set $\mathbf \Psi$. This set contains all the terms that enter into the matrices $J_I(\mathbf k)$ and $J_R$, for example $\Psi=\left(\epsilon_{12},J_{11},J_{12},\ldots\right)$
where $J_{ij}$ are entries of the Jacobian matrix $J_R$. Given the set of parameters $\mathbf \Psi$, we can reconstruct the Jacobians (i.e., the Jacobians are trivially functions of all their parameters $J_I\equiv J_I(\mathbf \Psi)$, and therefore whether the system is microphase or macrophase separated. We seek to understand how each of the members of the set $\mathbf \Psi$ affect the value of $P_J$ through the following decomposition:}
\begin{equation}\label{eq:PJIJC}
    P(\mathbf \Psi)=P_J(J_I(\mathbf \Psi),J_R(\mathbf \Psi))=  P_0 + \sum_{\psi \in \Psi}\Delta P_{\psi}(\psi) + \sum_{(\psi \in \Psi)}\sum_{(\theta \in \Psi, \psi \neq \theta)}\Delta P_{\psi,\theta}(\psi,\theta) + \hdots.
\end{equation}
\resub{where $\psi$ and $\theta$ is a single element of the set $\Psi$. This means that we express the full function $P$ as sums  of subfunctions that depend on zero parameters, one parameter, two parameters etc. For example, consider a generic function such as $f(y_1,y_2)=1+y_1-y_2 + y_1^2 y_2^2$: the subfunction of zero parameters would be $f_0=1$, the subfunction of one parameter would be $f_{y_1}(y_1)=y_1$ and $f_{y_2}(y_2)=-y_2$ and the subfuncion of two parameters would be $f_{y_1,y_2}(y_1,y_2)=y_1^2 y_2^2$, each of these terms therefore gives information about the action of a parameter independently or in concert on the value of the total function.}

 Each term in equation~\eqref{eq:PJIJC} represents how the probability to observe our feature of interest is modified through changing only a single parameter, two parameters concurrently etc. Due to the difficultly in analyzing higher order correlations, we focus on  two-point correlation functions and lower, and we determine the sub-functions $P_{\psi}$ through functional ANOVA~\cite{Hooker2004,Hooker2012}. While it is plausible that the design principles of the problem at hand are too complex to be represented in lower-dimensional form, we focus on this simplified case in order to identify easily discernible, pragmatic rules. 
 
 \resub{We computed the terms in Equation~\eqref{eq:PJIJC} using symbolic integration through \textit{Mathematica}, following~\cite{molnar2020interpretable}. The value of $P_0$ is the mean value of function $P$, and it can be interpreted as the probability of microphase separation computed over the set of random parameter choices; the value of each subfunction $\Delta P$ should be interpreted as how much the subfunction changes the value of function $P$ towards either 1 (microphase) or 0 (macrophase), away from $P_0$. We note the caveat that even though our function $P$ is either 1 or 0, the subfunctions need not be restricted to these values. To try to make this subfunction intuitive, imagine performing an experiment where one can only change one parameter (e.g. $\epsilon_{12}$), but each experiment is instantiated with random values of the other parameters. One sets e.g. $\epsilon_{12}=0.2$ and performs $M$ experiments, noting whether the system is microphase or macrophase separated. One repeats this for every value of $\epsilon_{12}$. By looking at the mean of the set of observations for each $\epsilon_{12}$ one is effectively computing the subfunction $\Delta P_{\epsilon_{12}}(\epsilon_{12})$, once the mean probability of observing microphase separation is subtracted. Higher order subfunctions of more than one variable are more involved, but the same intuitive picture should be kept in mind. We also note that as the measured values are either 0 or 1, the standard deviation and mean of a sequence should coincide.}
\subsubsection{Analytical example for two component system}
We begin to illustrate our approach using a two component example in which it is possible to determine analytically the conditions necessary for chemical dynamics to introduce spatial non-equilibrium structural features of the fluid. We can then compare analytical conditions on the parameters with the outcomes of a statistical evaluation of~\eqref{eq:PJIJC}. 

For a two-component system, the Jacobians of the spatial interactions and the chemistry are:
\begin{align*}
J_I(\mathbf k)=\begin{pmatrix}
 a |\mathbf k| ^2 - \gamma^2 |\mathbf k|^4 & -\mathcal{D} \epsilon_{12} |\mathbf k|^2\\
-\mathcal{D} \epsilon_{12} |\mathbf k|^2 & -\mathcal{D} |\mathbf k|^2
\end{pmatrix},\quad 
    J_R=\begin{pmatrix}
 J_{11} & J_{12}\\
J_{21} & J_{22}
\end{pmatrix},
\end{align*}
where $\epsilon_{12}$ are spatial couplings between species, $|\mathbf k|$ is the wave number, $\mathcal{D}$ is the diffusion coefficient, $\gamma$ is the surface tension, and $a$ is an energy scaling factor \resub{that incorporates the parameters involved in the Cahn-Hilliard equation such as $\nu, \rho_1,\rho_2$ etc.}
For simplicity, we set $a=1$ and $\mathcal{D}=1$. Different diffusion rates can lead to Turing patterns even in the absence of phase separation~\cite{Turing1952}, thus we exclude studying the effects of varying diffusion coefficients in order to focus our attention on the effects of chemical reactions. \resub{
This leads to a set of all the parameters given by:
$\mathbf \Psi = \left\{\epsilon_{12},\gamma,J_{11},J_{12},J_{21},J_{22}\right\}$
}

To study  the impact of $J_R$ on the \resub{dispersion relation}, we analyze the determinant of the sum of $J_I$ and $J_R$:
\begin{equation}\label{eq:jacobians}
\det(J_I(\mathbf k)+J_R) = \det(J_I(\mathbf k))+\det(J_R) +\det(J_I(\mathbf k))\text{Tr}(J_I(\mathbf k)^{-1} J_R ).  
\end{equation}

\resub{this is a helpful expression, as we know that the determinant itself is given by the product of the eigenvalues $\det(J_I(\mathbf k)+J_R) = \lambda_1 \lambda_2$. As we already specified, the determinant of the  Jacobian associated to the chemistry $J_R$ must be zero, and this leads to the following analytical expression:
\begin{equation} \label{eq:cyrves}
k^2 \left(J_{22} \left(1-k^2 \gamma^2 \right)+\epsilon_{12}
   \left(J_{12}+J_{21}\right)-J_{11}+k^2 \left(k^2 \gamma^2 -\epsilon
   _{12}^2-1\right)\right) = \lambda_1 \lambda_2
\end{equation}
}


In the case of a microphase separation, the determinant of $J_I(\mathbf k)+J_R$ must have at least 3 zeroes \resub{as a function of $|\mathbf k|$} (crossing points of the blue curve in Fig.~\ref{fig:fig0}). Thus, microphase separation corresponds to situations where ~\eqref{eq:cyrves} has three real zeros in $k\ge 0$. We can translate this requirement into conditions on the entries of $J_R$ and $\epsilon_{12}$, and are an example of the function $P$ defined earlier in~\eqref{DemixingRule}. Restricting our analysis to the range where $-1<\epsilon_{12}<1$, these conditions are:

\begin{minipage}{\textwidth} 
\begin{minipage}{0.4\textwidth}
$P(\Psi)=\begin{cases}
&\epsilon_{12} J_{12}+\epsilon_{12} J_{21}-J_{11}>0, \text{ and}\\
&\epsilon_{12} J_{12}+\epsilon_{12} J_{21}-J_{11}\leq \frac{\epsilon_{12}^4+2 \epsilon_{12}^2+1}{4 \gamma^2},\text{ and}\\ 
&-\epsilon_{12} J_{12}-\epsilon_{12} J_{21}+J_{11}-J_{22}<0,\text{ and} \\
&J_{2,2}<0,
\end{cases}$
\end{minipage} \qquad \qquad \textbf{or}
\begin{minipage}{0.5\textwidth}
\begin{equation} \label{eq:twospecies}
\begin{cases} 
&\epsilon_{12} J_{12}+\epsilon_{12} J_{21}-J_{11}>\frac{\epsilon_{12}^4+2
   \epsilon_{12}^2+1}{4 \gamma^2},\text{ and} \\
&\epsilon_{12} J_{12}+\epsilon_{12} J_{21}-J_{11}<\frac{\epsilon_{12}^2+1}{\gamma^2},\text{ and} \\
&-\epsilon_{12} J_{12}-\epsilon_{12} J_{21}+J_{11}-J_{22}<0, \text{ and}\\
&J_{22}<\frac{1-\epsilon_{12}^2}{\gamma^2 }-2 \sqrt{\frac{\gamma^2  \left(\epsilon_{12}   J_{12}+\epsilon_{12} J_{21}-J_{11}\right)-\epsilon_{12}^2}{\gamma ^4}}.
\end{cases}
\end{equation}
\end{minipage}
\end{minipage}

\vspace{4pt}

\resub{This function should be interpreted as a complicated logical sentence. Either all the clauses on the left hand side are true, \textit{or} all the clauses on the right hand side are true. If the sentence is true, the set of parameters corresponds to microphase separation, and if not it corresponds to macrophase separation. It is immediately apparent from this framing how complicated these expressions can become, even for 2 component systems. In fact, even the above expressions required computer algebra software (\textit{Mathematica}) to extract closed form expressions. Nevertheless, we can go through the process of describing design rules that arise in a two component system. }

While these expression are complex, the inequality $-\epsilon_{12} J_{12}-\epsilon_{12} J_{21}+J_{11}-J_{22}<0$ is simple enough that we can offer an interpretation of its meaning. For occurrence of microphase separation it is beneficial for term $J_{11}-J_{22}$ to be negative. In a negative definite matrix both  $J_{11}$ and $J_{22}$ are generally both negative, so it is desirable that $\vert J_{11} \vert > \vert J_{22} \vert$. This means that upon a perturbation, it is preferable for the concentration of species 1 to relax faster back to its equilibrium state than the concentration of species 2. In a two component system, we can understand this through the idea that components 1 and 2 both suppress their own concentration through reactions: given that there are only two species, the only way that this can occur is through the inter-conversion of the two species into one another. This condition implies more of species 1 is being converted to species 2 than vice versa, which is a necessary condition for the formation of a microphase separation as species 1 has the inbuilt tendency to phase separate, which has to be balanced by chemical flux (as has been observed before \cite{Zwicker2015}).

As for term $-\epsilon_{12} J_{12}-\epsilon_{12} J_{21}$, if the interaction between species 1 and 2 is repulsive ($\epsilon_{12}>0$), then it is preferable for species 1 and 2 to produce each other (positive off-diagonals); if it is attractive ($\epsilon_{12}<0)$, then it is preferable for species 1 and 2 to inhibit each other (such as by reacting together to form a different substrate). In other words, it is preferable for the chemical interactions to have an effect on the concentration that is opposite from the effect of the spatial interaction. This opposite effect is beneficial to maintaining microphase separation. \resub{While the remaining conditions are not easy to interpret, they can be computationally decomposed to reveal the interplay between terms.}

\resub{The conditions on the Jacobian elements in equations~\eqref{eq:twospecies}}  can guide the design of chemical reactions that promote the emergence of microphase separation in the presence of uncertainty or fluctuations of the spatial interaction parameters, making the structure of the chemical Jacobian more robust with respect to such uncertainties. For example, consider a system in which the following chemical reactions occur:
\begin{align}
    \ch{z_1 <>[ $k_a$ ][ $k_b$ ] z_2 } 
\end{align}
where $z_1$ is the phase separating species. The addition of the following reaction:
\begin{align}
    \ch{2 z_2 <>[ $k_c$ ][ $k_d$ ] z_1 + z_2 } 
\end{align}    
 would make it more likely for the overall Jacobian $J_I(\mathbf k)+J_R$ to satisfy the microphase separation conditions if  the spatial interactions between $z_1$ and $z_2$ were repulsive (where all the rates are positive but unknown), but less likely to do so if the interactions were attractive. 
     
\begin{figure}[h]
 \begin{center}
\includegraphics[width=180mm, frame]{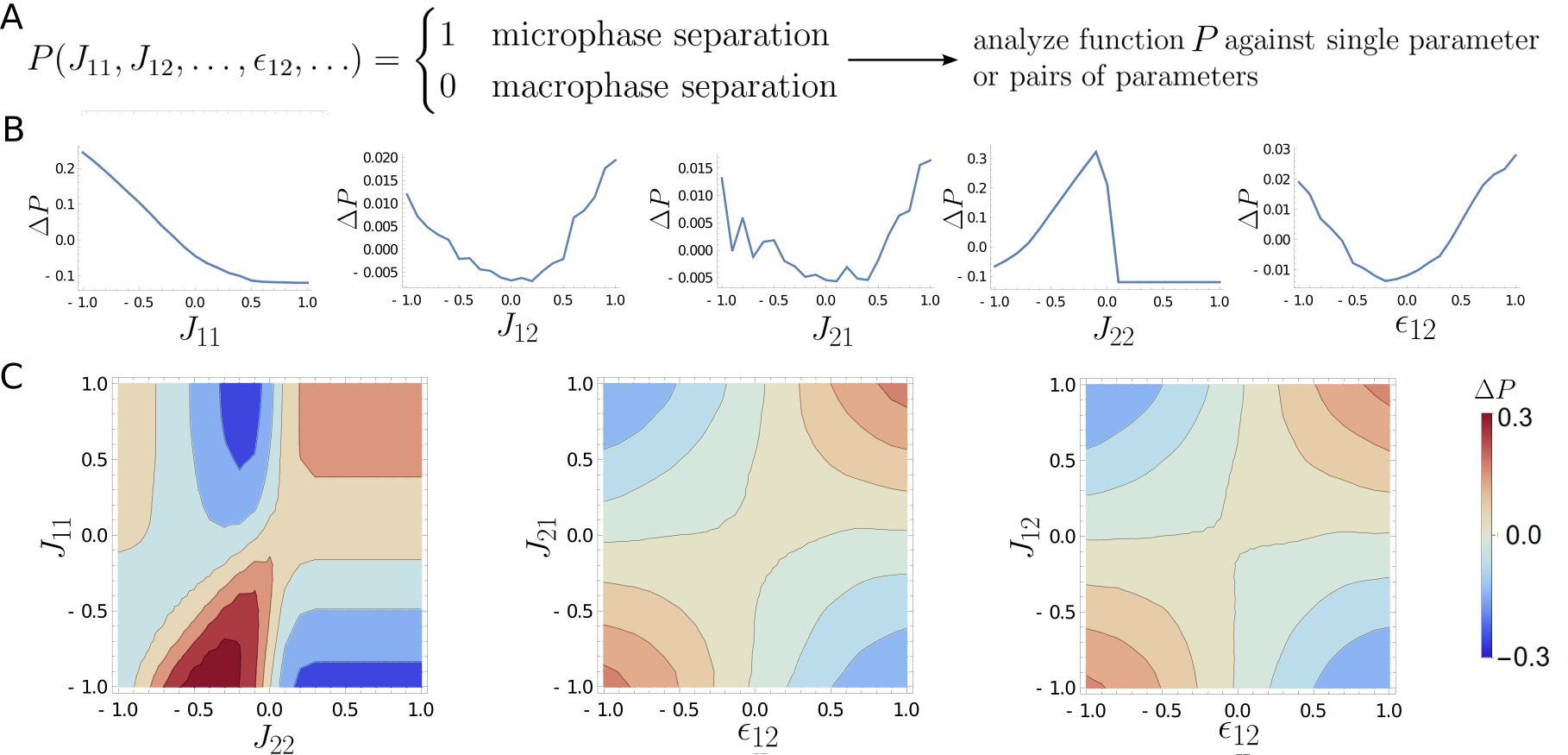}
\caption{How the various parameters of the Jacobian matrices impact the probability to observe microphase separation. \resub{(A) The function $P$ maps a set of parameters onto a binary yes-no function that tells us whether the system is microphase separated or macrophase separated.} Decompositions of this function are displayed as a function of both changing a single variable (B) and as a function of changing two variables simultaneously (C). Many such relationships could be generated, but we here show only those that have the largest impact on the function $P$. In particular, the interaction between $J_{12}$ and the spatial coupling $\epsilon_{12}$ is important, while components $J_{11}$ and $J_{22}$ are important in themselves. }
\label{fig:fig1} %
\end{center}
\end{figure}

\resub{The preceding discussion allows us to reason about relatively simple rules in the design of two component systems. These rules are rather intuitive in this case, yet it's tricky to generate them and evaluate them. Therefore, }next, we use \resub{decomposition method in the previous section} to illustrate how the likelihood of microphase separation is affected by changes in the parameters of the Jacobians in~\eqref{eq:jacobians}. Fig.~\ref{fig:fig1} shows the decompositions in which a selection of individual or of pairs of parameter are changed (the remaining sets are in Supplementary Fig.~1). \resub{Some of these relationships are consistent with those inferred} from~\eqref{eq:twospecies}, for example Fig.~\ref{fig:fig1}B confirms that the likelihood of microphase separation increases if $J_{11}$ is more negative than $J_{22}$. This means the deactivation of species 1 (the phase separating component) has to proceed most strongly, which is natural if we consider that phase separation causes a locally increase of concentration, a process that needs to be counteracted by chemical deactivation to avoid predominance of macroscopic separated phases. We also confirm our expectation that spatial interactions ($\epsilon_{ij}$) and chemical reactions ($J_{ij}$) between the two species should have opposite effects on the concentration.

\subsubsection{Random Jacobian analysis of systems with four components}

\begin{figure}[h]
\begin{center}
\includegraphics[width=180mm, frame]{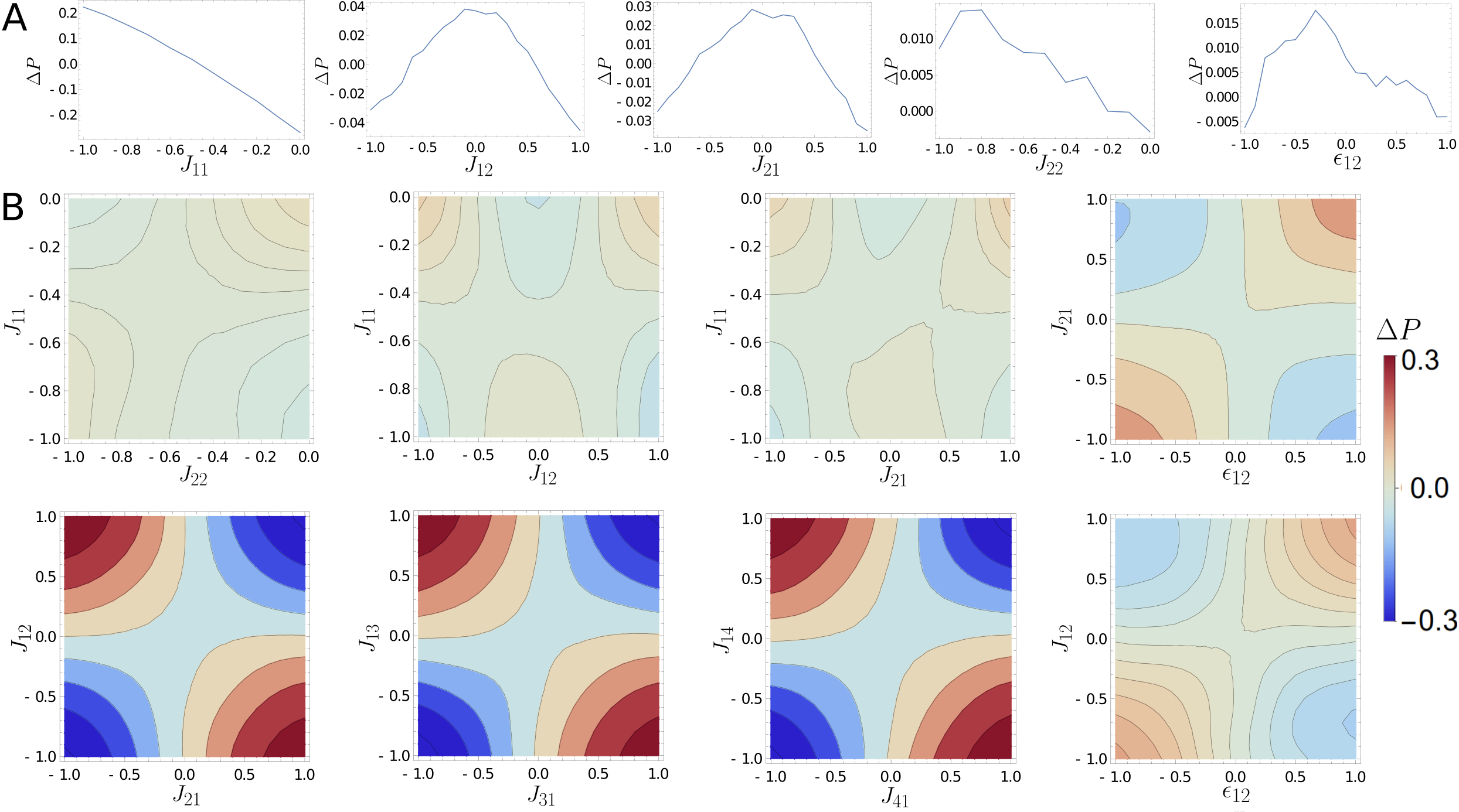}
\caption{Modification of the function $P$ observe a microphase separation as a function of various components of the Jacobians. For comparison, we show the complementary parameters as appeared in Fig. \ref{fig:fig1}. (A) the modification of the probability due to changing a single component and (B) the modification of the probability when two components are being modified simultaneously. While some features of the analysis remain for the case of 4 species, several qualitative differences emerge as to what confluence of parameters leads to high probabilities of observing microphase separation. }
\label{fig:fig2} %
\end{center}
\end{figure}

We use computational analysis to consider another example that includes four distinct species. For ease of comparison with the two component system, we restrict the spatial interactions between components to be between species $1$ and $2$, with all other components set to zero. In other words, only the components $\epsilon_{12}$ and $\epsilon_{21}$ in the interaction matrix of~\eqref{eq:CHF} are non-zero. Fig.~\ref{fig:fig2} shows how the likelihood of microphase separation changes as  parameters are varied. For ease of comparison with the two components system, we varied the same single parameters (top row panels in Fig.~\ref{fig:fig2}), identifying some similarities. For example, there is a strong dependence on the parameter $J_{11}$ where systems with more negative $J_{11}$ (strong self deactivation) \resub{are more likely to undergo microphase separation}. The other components, however, display differences to the two component example, the generic behavior of the other components seem to be inverted in comparison to the two component example, and the effect of each other component is much weaker (See Supplementary Fig.~2). 

The relation between $J_{12}$ and $\epsilon_{12}$ is similar to the two component example: {spatial and chemical interactions should have opposite effects on concentration to promote the emergence of condensates of finite size.} However, \resub{and in contrast to the observations made for the two component example}, we can now observe that there is very strong interactions between the complementary components of the first row and first column, i.e., $J_{12}$ and $J_{21}$. We can interpret this heat map by observing that the preference appears to be strongly in favor in terms of $J_{12}$ and $J_{21}$ being of different signs. \resub{Microphase separation is generally disrupted if the odd-diagonal terms are the same}.  This suggests that in order to design chemical networks leading to microphase separation it is necessary to produce a network where component 2 activates component 1 and component 1 inhibits component 2 or \textit{vice versa}. This corresponds to instability in the level of components 1 and 2 when they are mixed together, i.e., the creation of fluxes, which would always be a precondition of fixed size droplets. {We observe a similar pattern when considering the chemical interactions between species 1 and 3, and 1 and 4.} \resub{Analysis of figure  Fig.~\ref{fig:fig2} furthermore reveals an element of frustration\cite{Chowdhury1986} in the problem. As $J_{12}$ and $J_{21}$ should have \textit{opposite} signs, and yet each of them as they appear in the plots of $J_{12} \text{ vs } \epsilon_{12}$ and $J_{21} \text{ vs } \epsilon_{12}$ would prefer to have an $\epsilon_{12}$ which \textit{matches} their own sign. It's not possible to satisfy this constraint, which may explain why $\epsilon_{12}\ne0$ is relatively less important for systems above more than two components. In general, we believe that increasing the number of components in this system increases the chances that frustration such as this occurs, suggesting an intriguing problem of optimization in parameter space for design of robust networks.}

\subsection{Identifying chemical reaction networks that achieve condensate size control}
\begin{figure}[h]
\begin{center}
\includegraphics[width=180mm, frame]{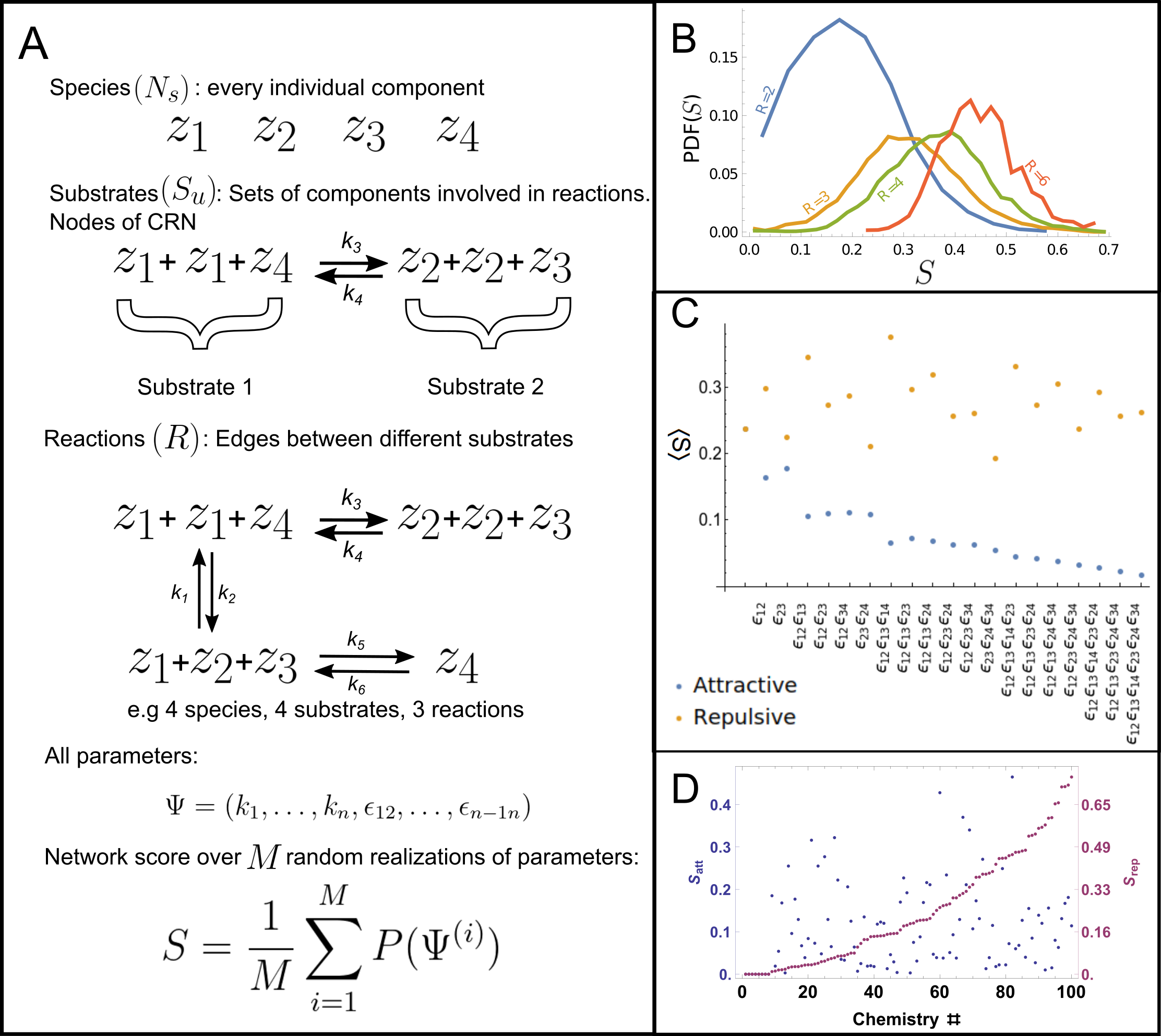}
\caption{{Comparing relative efficacy of different chemical reaction networks and spatial interactions to produce a system with a \resub{microphase separation}. (A) Schematic of the CRN scoring approach. We begin with a system with a set number of species, and then randomly generate $S_u$ substrates and $R$ reactions between them. We then interrogate how often the feature we are interested in arises across random choices of all the parameters $\Psi$. (B) The average score changes depending on the number of reactions across different CRNs; here we assume $N_s=4,S_u=4$. More reactions have a positive impact on $\mathcal{S}$.  (C) Impact of spatial interactions on the average score across different chemistries for purely repulsive or attractive interactions; each of the x-axis labels is the subset of spatial couplings that is non-zero. (D) Scores against 100 distinct labelled chemistries when the interactions are changed to be positive or negative. The same chemistries which do well for repulsive interactions do not correlate with the score for attractive interactions.}} 
\label{fig:fig3} %
\end{center}
\end{figure}

The prior sections expounded upon the mathematical features necessary towards the realization of chemical reactions that produce droplets of finite size, which was conducted through analysis of the properties of the Jacobians arising from the interplay of spatial and chemical interactions among species. \resub{This leads to generic rules on the level of activation/inhibition in the Jacobian, however, it is usually not possible to design Jacobians directly, rather, we are more used to introducing specific chemical reactions into a system.  In this section, we seek \textit{design rules for chemical reactions}  that produce microphase separation, following the workflow illustrated in Fig.~\ref{fig:fig3}A.}

\resub{We consider general chemical reaction networks (CRN) that can be described by a tuple $(N_s,S_u,R)$, where $N_s$ is the number of species, $S_u$ is the number of substrates (where a substrate is a set of reactants or products appearing in the reactions), and $R$ is the total number of reactions~\cite{VanDerSchaft2011}. Given a particular set of reactions, a CRN can be associated to an incidence graph whose nodes correspond to the substrates, and arcs corresponds to the reactions inter-converting the substrates. (Note that many different graphs can be associated with a particular tuple $(N_s,S_u,R)$, depending on the reactions chosen). Each graph admits a multitude of parameter realizations (reaction rate parameters). Once reactions and parameters are specified,  we can generate the time evolution equations of each CRN via the law of mass action. Now the entries $J_{ij}$ of the chemical reaction Jacobian matrix $J_R$  are specified by physical reaction rates and concentrations that populate the parameter vector $\Psi$. To test whether a CRN introduces microphase separation, we determine the algebraic form of its Jacobian $J_R$, and analyze the resulting dispersion relation that includes algebraic expression with random parameters. Our automated process for defining and examining random CRNs takes advantage of definitions in~\cite{VanDerSchaft2011}, and  is described in Section 1 of the Supplementary Note.}

An actual chemistry being a symbolic representation of interconversions between species, e.g. $\ce{z_1 + z_2 <=> z_3},~\ce{z_1 + z_4 <=> z_2} $ etc., what can we say about the \textit{structural} features of CRNs that make size control more or less likely? There is a degree of arbitrariness to this problem, in that the structure of the CRN (defined by the number of species $N_s$, the number of substrates $S_u$, the number of reactions $R$ in the system,  and the substrate/reaction graph) is not sufficient to define whether the Jacobians will yield microphase or macrophase separation, which is influenced by a myriad of parameters involved, such as the reaction rates, diffusion coefficients, surface tensions, interaction parameters etc. However, broadly stated, we can define this problem as one of optimization, introducing the following score metric:
\begin{equation} \label{eq:metric}
    \mathcal{S} = \left(\int \rho(\mathbf \Psi)P_J(J_I (\mathbf \Psi),J_R(\mathbf \Psi) ) \mathrm d \mathbf \Psi \right)/ \left(\int \mathrm d \mathbf \Psi \right)
\end{equation}
where we use the function $P_J$ defined in~\eqref{DemixingRule} that maps the Jacobian matrices to a dispersion relation (with $\text{microphase separation}=1$ and $\text{macrophase separation}=0$). The function $\rho(\mathbf \Psi)$ defines some probability distribution of parameters in the system. Maximization of $\mathcal{S}$ leads to a system which displays microphase separation over many different choices of parameters. Such a system would therefore be ``robust'' to uncertainty in parameters towards achieving size control. This optimization problem will also make it possible to identify patterns in the Jacobian $J_R$ (non zero entries and their sign) that maximize the number $\mathcal{S}$ over all parameter values\resub{, and we those patterns are consistent with those identified via the reaction-agnostic approach described in Section B}. We choose $\rho(\mathbf \Psi)=1$ because the probability distribution of parameters, which would depend on experimental conditions, uncertainty and temporal fluctuations etc., is not known in principle.

We use $\mathcal{S}$ in~\eqref{eq:metric}  as a metric to compute a ``size control score'' and rank the capacity of various CRNs to yield microphase separation (Fig.~\ref{fig:fig3}A). We generated CRNs by considering all possible combinations of $N_s=4$, $S_u\in[2,4]$, and $R\in [2,6]$. For each combination of $(N_s,S_u,R)$ we generated a series of graphs; for each interaction graph, we generated 1000 random parametric realizations.  (Further details about generation of random chemical reaction networks are in the Supplementary Note, Section 1.)  For physically meaningful results, as done earlier, we restricted our attention to closed networks conserving the total mass. We evaluated $\mathcal{S}$ using Monte Carlo integration by  obtaining a score (1 or 0) for each set of parameters, and we computed an average score for each CRN (for this purpose, we used the fully connected interaction matrix $J_I$). Fig.~\ref{fig:fig3}B shows the distribution of scores over networks with the same number of reactions. Interestingly, the more fully connected the network structure is, the more likely it is to generate microphase separation. In contrast, we found that CRNs with only two chemical reactions ($R=2$) have the lowest values of score $\mathcal{S}$, as shown in Fig.~\ref{fig:fig3}B. \resub{Consideration of the analysis of the Jacobian $J_R$ in the previous section leads to a plausible mathematical hypothesis regarding why, as with larger numbers of reactions the entries in the Jacobian $J_R$ become less correlated and can also take values with opposite signs. Highly correlated entries in the Jacobian could constrain the off-diagonal elements in such a way as to be destructive to the existence of microphase separation, as illustrated in Fig.~\ref{fig:fig2}.}

\begin{table}
\centering
\begin{tabular}{|c|c|}
\hline
{\bf $\quad$ Chemical reactions $\quad$ } & {$\quad$ \bf Score $\mathcal{S}$ $\quad$}\\
\hline 
$z_1+z_2\xrightleftharpoons{} z_4$& $\quad 0.596 \quad $\\
 $2z_1\xrightleftharpoons{} z_2$  & \\ 
\hline
$z_1+2z_3 \xrightleftharpoons{} z_2$ & $\quad 0.595 \quad$ \\ 
$2z_1 \xrightleftharpoons{} z_3$ & \\ 
\hline
$z_1+2z_2 \xrightleftharpoons{} z_3+2z_4$ & $\quad 0.591 \quad $ \\
$2z_1+z_4 \xrightleftharpoons{} z_2$ & \\
\hline
$z_1+z_3+z_4 \xrightleftharpoons{} z_2$ & $\quad 0.583 \quad $ \\
$2z_1 \xrightleftharpoons{}  z_4$ & \\ 
\hline
\end{tabular}
  \caption{Example highly scoring chemistries for $N_s=4,S_u=4,R=2$}
  \label{tbl:excel-table}
\end{table}

\resub{We show the networks with the highest score in Table \ref{tbl:excel-table}. By observing this table, we note that the best performing CRNs in terms of achieving microphase separation include the following class of reactions:}
\begin{align}\label{eq:ce1}
z_1  &\xrightleftharpoons[k_2]{k_1} z_i \\ \label{eq:ce2}
 z_1 + z_i&\xrightleftharpoons[k_4]{k_3} z_j 
\end{align}
where we can use $i,j$ ($i\not = j$) interchangeably for any of the non-phase separating chemicals in our system.  The CRN should include a reaction that converts $z_1$  to some other chemical $z_i$, and in addition this chemical $z_i$ can itself sequester $z_1$, thereby generating another chemical $z_j$. As a consequence, a local increase in $z_1$ is counterbalanced by sequestration of $z_1$ itself, which is qualitatively equivalent to self-repression and generates a  negative feedback loop. In other words,  size control is promoted by the presence of chemical reactions introducing a self-regulation mechanism in the local level of $z_1$. This is the best performing motif in a system where the spatial interactions can take random values.

\resub{ We can use the Jacobian analysis of the previous section to understand why the form of CRN shown in equations~\eqref{eq:ce1} and~\eqref{eq:ce2} is so good for microphase separation, (using $i=2,j=3$), the Jacobian corresponding to this CRN is given by:
\begin{equation}
J_R=\left(
\begin{array}{cccc}
 -k_4 \left[z_2\right]-k_2 & k_1-\frac{k_4 \left(-\frac{k_1 k_4 \left[z_2\right]^2}{k_2}-k_1 \left[z_2\right]\right)}{-k_4 \left[z_2\right]-k_2} & k_3 & 0 \\
 k_2-k_4 \left[z_2\right] & -\frac{k_4 \left(-\frac{k_1 k_4 \left[z_2\right]^2}{k_2}-k_1 \left[z_2\right]\right)}{-k_4 \left[z_2\right]-k_2}-k_1 & k_3 & 0 \\
 k_4 \left[z_2\right] & \frac{k_4 \left(-\frac{k_1 k_4 \left[z_2\right]^2}{k_2}-k_1 \left[z_2\right]\right)}{-k_4 \left[z_2\right]-k_2} & -k_3 & 0 \\
 0 & 0 & 0 & 0 \\
\end{array}
\right)
\end{equation}
where we consider the equilibrium concentration $\left[z_2\right]$ to be a free parameter . We see that the \textit{structure} of the CRN leads to a Jacobian where the off-diagonal entries at $\{1,2\},\{2,1\},\{1,3\},\{3,1\}$, which are most important towards the realization of microphase separation, depend on largely different parameters. Therefore, even when these parameters cannot be controlled, they will be more likely take different values, which is useful for size control as can be seen from our map in Fig.~\ref{fig:fig2}. This is related to the results for the more highly connected networks being better for microphase separation, as small CRNs tend to produce off-diagonals that are positively correlated, or off diagonals that can only take positive values , which as was seen in the previous section, is an impediment to realizing size control.
}

We are left with the question of what is the impact of the spatial coupling terms $\epsilon_{ij}$ on the size control score $\mathcal{S}$. Fig.~\ref{fig:fig3}C show the average value of $\mathcal{S}$ when only a subset of $\epsilon_{ij}$ couplings have a finite value (as specified in the x-axis), while all others are set to zero. Several trends can be observed. The best performing system is the one in which component 1 has a repulsive interaction to all the other species $2,3,4$. It can also be seen that the more attractive interactions the system has, the more difficult it is to realize a microphase separation (low average scores). A simple intuitive picture that would explain this is that components tend to get concentrated together when there are attractive interactions, and this makes it difficult to achieve a flux out of the dense phase that is necessary to arrest coarsening. 

\resub{To examine to what extent a CRN is likely to achieve size control under different types of spatial interactions,  we compare the scores of the same CRN against repulsive and attractive spatial interactions in Fig.~\ref{fig:fig3}D. In other words, the chemistry is the same, and all that is changed is all the entries of $\boldsymbol \epsilon$ are either all repulsive or attractive.  One way to test this is by generating the average score for 100 distinct CRNs for repulsive interations, and plotting them in a rank-ordered manner. These same CRNs can be used to compute the scores for attractive interactions. If the curves were to display similar trends, this would suggest that a CRN that is good in a system with repulsive interactions also tends to be good in a system with attractive interactions. Thefore The lack of correlation between the two curves in Fig.~\ref{fig:fig3}D suggests a well performing chemistry in the presence of repulsive spatial interactions will not necessarily be good when the interactions are attractive. Therefore, the problem of designing CRNs for size controlled condensates is quite sensitive to the sign of the spatial interactions in the system. }

\subsection*{Controlling dynamics of condensates through chemical reactions}

The approach we have described can be adapted for different conditions. One interesting extension is the emergence of dynamical phenomena. This cannot be exactly enumerated, but a condition we can obtain from the dispersion relation $\beta(|\mathbf k|)$ is:
\begin{equation} \label{eq:Dyn}
    \Re(\beta(|\mathbf k|) = 0 \text{ and } \Im(\beta(|\mathbf k|))>0
\end{equation}
which would correspond to an oscillations. Obviously, in the real system, we have many different wavelengths, each of which has its own real and imaginary value of the eigenvalues of the Jacobian matrix, and therefore~\eqref{eq:Dyn} is not a sufficient condition for the realization of sustained oscillations in our system. However, it is still of interest to analyze what kinds of patterns in the Jacobian elements could possibly lead to oscillations.
\resub{To that end we define a new function:
\begin{equation}\label{eq:OscilRule}
    Q_J(J_I(\mathbf k),J_R)=
\begin{cases}
1 &\text{if an instability with complex eigenvalues exists}\\
0 &\text{otherwise}
\end{cases}
\end{equation}
}
\resub{In words, we look for dispersion relations where both $\Re(\beta(|\mathbf k|)>0$ and $\Im(\beta(|\mathbf k|))>0$ exists for some value $\mathbf k$}

\begin{figure}[h]
\begin{center}
\includegraphics[width=180mm, frame]{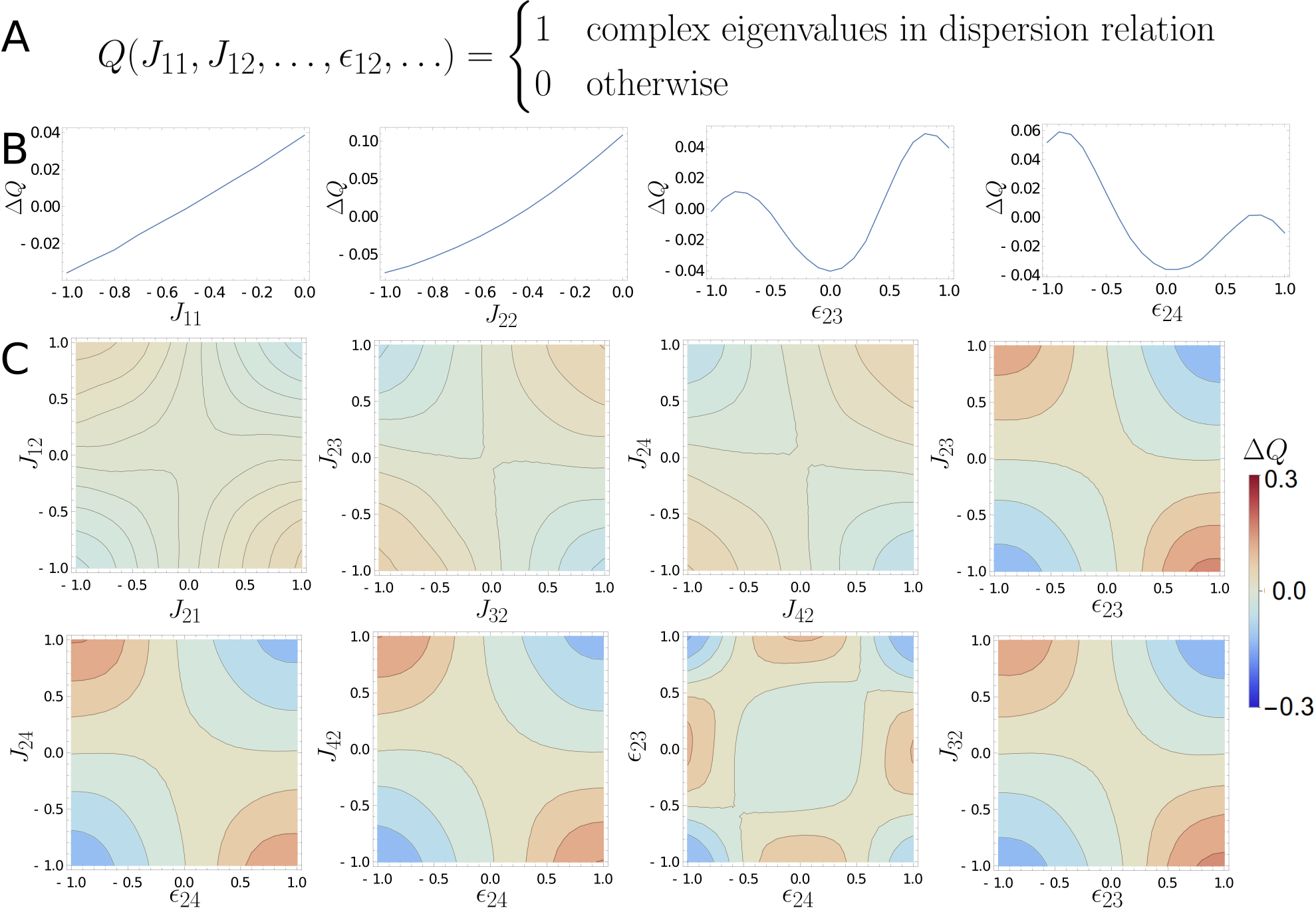}
\caption{Function describing the presence of a dispersion relatioDynamicallyn with both imaginary and real components, see~\eqref{eq:OscilRule}, with arguments as the Jacobians. Presence of imaginary eigenvalues would be indicative of long lived dynamical states. Only the features which have the strongest impact on the probability are shown, both for varying a single parameter (B) and for varying two different parameters simulataneously (C). In contrast to the observations for \resub{microphase separation}, different sets of components are most important to the observation of imaginary eigenvalues, suggesting the possibility that these two different features needn't interfere with one another. Unlike size control, the presence continued dynamics depends more strongly on the details of the spatial interactions, which means that chemical motifs are of relatively less importance.}
\label{fig:fig4} %
\end{center}
\end{figure}

\resub{We can do the same decomposition we did searching for microphase separation} for a four species system, shown in Fig.~\ref{fig:fig4}. Patterns similar to those in Figs. ~\ref{fig:fig1} and~\ref{fig:fig2} emerge in the relationships between field couplings $\epsilon_{ij}$ and Jacobian elements $J_{ij}$.  However, the variables most of interest in the Jacobian towards oscillatory behaviors are the components that were relatively unimportant toward \resub{microphase separation}, such as the coupling between components $2$ and $3$ (which aren't the components that phase separate). The Jacobian elements themselves display a coupling feature where it is advantageous for both components to have the same magnitude. Altogether, this would mean that the best way towards achieving dynamical oscillations of droplets would be to couple the droplets to another reaction-diffusion system which already displays oscillatory behavior, rather than attempting to induce dynamical oscillations of the droplets themselves.

As is well understood, linear stability analysis is not fully predictive of the properties of the long time behavior of reaction diffusion systems, nor can it precisely enumerate exactly how the long time behavior will ``look''~\cite{Riaz2007}. In order to fully understand what these systems do, there are as of yet few alternatives towards complete solution of the equations involved. Given the thousands of different possible CRNs, full enumeration of the possibilities here are prohibitive, but we can take a taxonomic approach towards the problem by simulating a few examples with interesting dispersion relations and observing the time evolution. The subset we are interested in here shall be those dispersion relations which display both microphase separation and dynamic oscillations. These can be generated by generating many CRNs and finding the parameter sets that display both of these features. The taxonomy is displayed in Fig.~\ref{fig:fig5} (the corresponding reactions are in the Supplementary Note Section 3). Microphase separation arises rather commonly in our system, though the precise structure of this phase itself displays plentiful variance, including droplets and stripes (as is well established). \resub{Whether the exact features of the final states displayed here depend on ``thermodynamic" or ``chemical" factors would require further analysis of individual reaction schemes. However, we show in Section 2 of the Supplementary Note that in the absence of chemical reactions none of the observed states could occur, as such systems will only display macrophase separation. Chemical reactions are therefore necessary towards the achievement of size limited patterns, but the precise form these patterns take will depend on the interplay of the conserved and non-conserved dynamics. } 

When looking for dynamics, the linear stability analysis isn't strongly predictive of long lasting dynamic oscillations. In fact, most chemistries we attempted lead to slow exponential relaxation to a steady state as opposed to long lasting oscillations, despite the presence of complex eigenvalues. From this we would surmise that the conditions required to obtain long lasting oscillatory dynamics would require some additional analysis accounting for the non-linearities. Another possibility is that the presence of a noise term in our system may lead to excitation of dynamical modes, but we have foregone the study of noise in this paper. 

\begin{figure}[h]
\begin{center}
\includegraphics[width=180mm, frame]{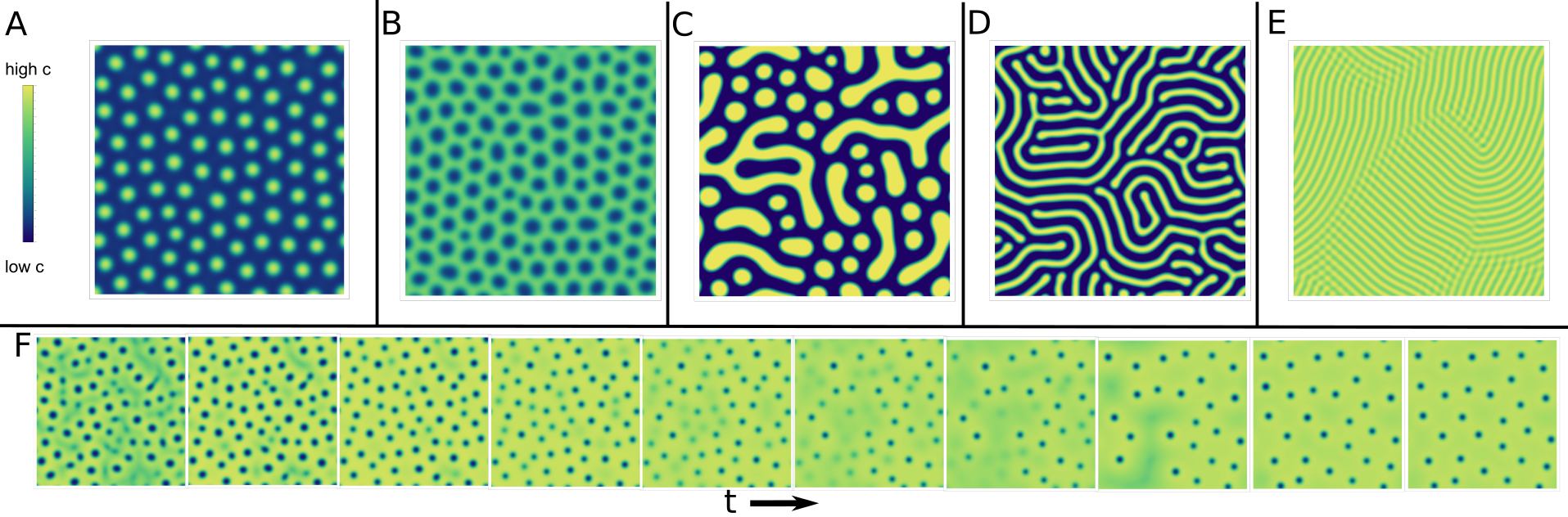}
\caption{Examples illustrating the taxonomy of possible (real) solutions (concentration of the phase separating material) found through the condition~\eqref{eq:Dyn}. These solutions display significant visual differences, although they all exhibit \resub{microphase separation}. Under these schemes, we observe (A) droplets of finite size (B) anti-droplets (droplets of low density) (C) Irregular droplets (D) Stripes (E) Stripes of Larger width. (F) Condensation dynamics introduced by a particular CRN. The dynamics shows droplets coming into and out of existence, however, the concentration field eventually settles into a static state. The CRNs corresponding to each example are in the Supplementary Note, Section 3.}
\label{fig:fig5} %
\end{center}
\end{figure}

\section{Discussion and Conclusion}

While biomolecular phase separation can be controlled over time by modulating temperature and ionic conditions~\cite{dignon2019temperature}, changes in these properties are often limited by the necessity to maintain conditions compatible with life. This obstacle can be circumvented by controlling phase separation through a change of subunit concentration, which is easily achieved via CRNs that allow for interconversion between species~\cite{kirschbaum2021controlling}. We have discussed how the coupling of phase separating molecules with chemical reactions can lead to the emergence of non-equilibrium behaviors, such as the existence of droplets with size showing a well defined length scale or  with size that undergoes continuous temporal dynamics. Through  computational simulations, we have identified patterns on the interaction parameters and we scored CRN structures that support the realization of these features. 
We have not considered the effects of either momentum transport, or random noise, and we acknowledge that both of these elements could conceivably alter some of the relationships we have observed, though we expect our broad findings to hold.

Stabilization of condensate size in a binary mixture has been computationally demonstrated before through the introduction of externally provided fuel and waste species, or through an enzyme that self-segregates into the droplets generating diffusive fluxes~\cite{kirschbaum2021controlling}. Coarse grained simulations of individual molecules in a system that contains phase separating species showed that a particular class of chemical reactions proceeding at a given rate can lead to novel non-equilibrium behaviors, including molecular sorting, spatial and temporal oscillations and tunability of chemical production~\cite{Osmanovic2019}. However, exploring chemical rules that lead to  non-equilibrium behavior cannot be done efficiently via coarse grained molecular simulation. Capitalizing on these studies, our work considers a broad set of CRNs with the goal of identifying useful design rules.

\begin{SCfigure}[\sidecaptionrelwidth][t]
\includegraphics[width=45mm, frame]{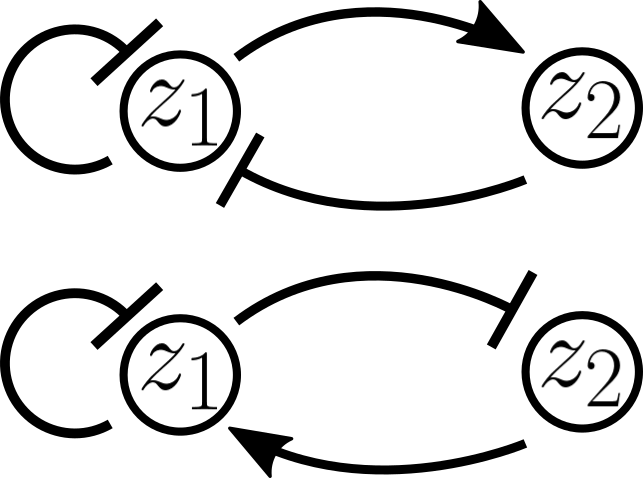}
\caption{\resub{Two different kinds of motifs in terms of generalized chemical reactions that are most helpful towards the realization of microphase separation. Component 1 (the phase separator) should self-inhibit, and either inhibit another component $2$ that activates $1$, or activate a component $2$ that inhibits $1$} }
\label{fig:figact} %
\end{SCfigure}

By framing the problem in terms of the properties of the Jacobians, we were able to identify parameter trends that should be satisfied to achieve particular behaviors. For \resub{microphase separation} to occur (size control), we found that a) chemical reactions should deactivate the phase separating species, by converting it into a species that does not participate in condensation and rather diffuses in the system; and b) the impact of other chemicals on the phase separating species should have opposite effects relative to the impact on them of the phase separating species. \resub{For example, if the phase separating molecule is species 1, and species $2$ is another molecule, we found that either 1 should produce $2$ and $2$ should inhibit 1, or that $2$ should produce 1 and 1 should inhibit $2$, and the system is at rest only when the level of 1 or $2$ is zero. This shows that size control benefits from the presence of chemistry that makes it unlikely for species $1$ and $2$ to coexist in the same region of space, and may indicate the presence of a negative feedback loop. We illustrate this idea in Fig.~\ref{fig:figact} using a network diagram often used in biology to represent activator/inhibitor interactions~\cite{alon2006introduction}.} When exploring the capacity of randomly generated chemical reaction networks to generate \resub{microphase separation} by analyzing their Jacobian, we found a recurring CRN motif in which the deactivated phase separating molecules further deactivate other active molecules via molecular sequestration, which has been associated with perfect adaptation in biology~\cite{briat2016antithetic,samaniego2021ultrasensitive}.  This reaction introduces negative feedback, because the higher the level of phase separating species, the more self-repression occurs.  This leads to large fluxes of material out of droplets, counteracting the influx of phase separating species. Yet, the existence of negative feedback does not represent a necessary nor sufficient condition towards achieving \resub{microphase separation}, because all the patterns we have described here are only statistical in nature. This finding is still interesting because negative feedback has a well-known  role in the context of stabilization and emergence of oscillations in biomolecular networks~\cite{blanchini2014structural,briat2016antithetic}. \resub{Such relationships only arise when considering more than two components, demonstrating the delicate nature of trying to understand the behavior of many coupled fluids.}

Curiously, we  observed that the more coupled the CRN is, the better it performs in terms of size control. It appears as a general effect, therefore, that the more chemical fluxes are included in the system, the more likely is the realization of \resub{microphase separation}, even if individual fluxes might not be optimized towards that actual goal. This would suggest that the presence of any random chemical flux corresponding to some energy-consuming reaction in a phase separating system rarely has the effect of stabilizing the macroscopically separated state, rather, it would appear to be the opposite. \resub{A mathematical explanation for this phenomenon can be gained from the full analysis of the function $P$ }

We also found that spatial interaction parameters have a major influence on the design of CRNs achieving \resub{microphase separation}. We found that repulsive couplings, i.e., the different species like to spatially segregate, are generally beneficial. In contrast, attractive couplings make it more difficult to design CRNs for size control. This is consistent with the common sense notion that it is harder to regulate the size of condensates were all the spatial interactions in the system attractive, for this would tend to promote the aggregation of all objects, to the detriment of control over their size. This finding is important in the context of experimentally designing molecular subunits, for example using peptides or nucleic acids, that can present non-specific, undesired attractive interactions often classified as ``cross-talk''~\cite{zhang2011dynamic}. These interactions can perturb the system's parameters enough to suppress its intended behavior, an effect that is particularly critical in non-equilibrium molecular systems~\cite{green2019autonomous}.  In the case of \resub{microphase separation}, the presence of any kind of spatial attractive interaction would make it much harder to realize that goal. Indeed the design of a CRN that doesn't take this into account could mean the theoretical result would be hard to reproduce. A way around this problem would be to include the unknown parameters as a constraint, and identify the CRN forms that lead to the best performance \textit{given} that attractive spatial interactions exist. While here we explored CRNs assuming that all the chemicals are coupled randomly, our analysis can be repeated including experimental limitations and  constraints like the presence of unwanted attractive couplings. 

When considering the emergence of periodic behaviors over time in our phase separating system, we noted that couplings between all the non-phase separating components appear to be most important (for example mediated by the interaction between species 2 and 3, and 2 and 4). We speculate that having stable oscillations in the phase separating material itself is difficult due to the fact that aggregation compromises the conditions required for the presence of sustained oscillations. In this case, the other species in the system may oscillate when decoupled from the phase separating species, and oscillations observed in the droplets may be a secondary effect of oscillations of chemicals in the dispersed phase. In practice, it is rare to observe stable oscillations in this system when the full partial differential equations were solved, though we have observed systems with {limited} oscillations of droplets popping into and out of existence.

A major advantage of using CRNs to control separation, when compared to introducing global environmental changes in temperature and solvent, is that CRNs could direct independently numerous coexisting condensates. The analysis of biological CRNs will help formulate hypotheses on how  cells manage a multitude of types of condensates  with distinct composition, function, structure, and temporal dynamics~\cite{kirschbaum2021controlling,Zwicker2015,shin2017liquid}. Our results will also provide guidance toward the design of novel materials in the context of a rapidly expanding set of designeable molecular substrates that could be used to implement a variety of CRNs~\cite{chen2013programmable,chen2021programmable}, as well as a variety of condensates~\cite{kaur2021sequence,gong2022computational,mountain2019formation}. \resub{We have used the general model presented here to gain insight on the behavior of artificial DNA condensates in which subunits are activated and deactivated through chemical reactions, and we found  experiments to be consistent with theoretical predictions~\cite{agarwal2022biochemical}. We thus expect that our approach will be useful to explore whether chemical reactions could provide instructions to manage condensate formation, dissolution, organization, and other macroscopic properties.}

\acknowledgements{EF and DO acknowledge support from NSF FMRG:Bio award 2134772 to EF and from the Sloan Foundation through award G-2021-16831.}


\bibliography{size_control_references}

\end{document}